\documentclass[twocolumn,aps,prc,floatfix]{revtex4}
\usepackage{graphicx}
\usepackage{dcolumn}
\usepackage{bm}

\begin{document}

\title{The interplay of short-range correlations and nuclear symmetry energy in hard photon productions from heavy-ion reactions at Fermi energies}

\author{Gao-Chan Yong$^{1}$}\email{yonggaochan@impcas.ac.cn}
\author{Bao-An Li$^{2}$}\email{Bao-An.Li@tamuc.edu}

\affiliation{%
$^1${Institute of Modern Physics, Chinese Academy of Sciences, Lanzhou
730000, China}\\
$^2${Department of Physics and Astronomy, Texas A$\&$M
University-Commerce, Commerce, TX 75429-3011, USA}
}%

\begin{abstract}
Within an isospin- and momentum-dependent transport model for nuclear reactions at intermediate energies, we investigate the interplay of the nucleon-nucleon short-range correlations (SRC) and nuclear symmetry energy
$E_{sym}(\rho)$ on hard photon spectra in collisions of several Ca isotopes on $^{112}$Sn and $^{124}$Sn targets at a beam energy of 45 MeV/nucleon. It is found that over the whole spectra of hard photons studied,
effects of the SRC overwhelm those due to the $E_{sym}(\rho)$. The energetic photons come mostly from the high-momentum tails (HMT) of single-nucleon momentum distributions in the target and projectile.
Within the neutron-proton dominance model of SRC based on the consideration that the tensor force acts mostly in the isosinglet and spin-triplet nucleon-nucleon interaction channel, there are equal numbers of neutrons and protons, thus a zero isospin-asymmetry in the HMTs. Therefore, experimental measurements of the energetic photons from heavy-ion collisions at Fermi energies have the great potential to help us better understand the nature of SRC without any appreciable influence by the uncertain $E_{sym}(\rho)$. These measurements will be complementary to but also have some advantages over the ongoing and planned experiments using hadronic messengers from reactions induced by high-energy electrons or protons. Since the underlying physics of SRC  and $E_{sym}(\rho)$ are closely correlated, a better understanding of the SRC will in turn help constrain the nuclear symmetry energy more precisely in a broad density range.
\end{abstract}

\maketitle

\section{Introduction}

The study on nucleon-nucleon short-range correlations (SRC) in nuclei and nuclear matter has a long and fruitful history, see, e.g., refs. \cite{bethe71,anto88,Ben93,Pan99,sci08,Arr12,sci14,Claudio15,Ryc15,RMP2017} for reviews.
The growing interests in the SRC physics far beyond its traditional field and new efforts are strongly motivated by its fundamental importance for both nuclear physics and astrophysics. The continuous efforts have been powered by new discoveries in a series of experiments using proton-nucleus, electron-nucleus and photon-nucleus reactions over many years. For example, proton-removal experiments using high-energy electron or proton beams showed that about 20\% nucleons in medium-heavy nuclei are correlated \cite{sci08,e93,e96,sci14} due to the short-range tensor interactions predominantly in the isosinglet and spin-triplet neutron-proton pairs \cite{tenf05,tenf07}. The SRC pairs have large relative momenta but small center-of-mass (CMS) momenta \cite{pia06,sh07}. They lead to a high-momentum tail (HMT) (and simultaneously a depletion ) in the single-nucleon momentum distribution above (below) the Fermi surface \cite{bethe71,anto88,Claudio15,Rios09,yin13}. Moreover, the shape of the HMT is almost identical for all nuclei from deuteron to very heavier nuclei \cite{Ciofi96,Fantoni84,Pieper92,egiyan03}. Furthermore, extensive theoretical and experimental studies indicate that the HMT varies approximately with momentum $k$ according to $1/k^{4}$ \cite{hen14,sci14,henprc15,liba15}.
Very interestingly, the size of the HMT was found strongly isospin dependent. More quantitatively, the number of neutron-proton SRC pairs was found to be  about 18 times that of proton-proton or neutron-neutron pairs \cite{sci08,sci14}.

Despite of the interesting discoveries and extensive studies made so far, there are still many unresolved issues regarding the nature and impact of SRC especially in dense neutron-rich systems, such as, its mass and isospin dependence as well as the role of three-body correlations, see, e.g., refs. \cite{mosel2016,claudio2017}. For example, properties of the short-range tensor force and ramifications of the HMT on the Equation of State (EOS) of neutron-rich matter \cite{caili2017,caili4,yongp3,yongp4,yongprc2016}, especially the density dependence of nuclear symmetry energy $E_{sym}(\rho)$ are still poorly known. The $E_{sym}(\rho)$ encodes information about the isospin dependence of nuclear EOS, it is measured in the so-called parabolic approximation of the EOS by the change in single nucleonic energy when all protons are replaced by neutrons \cite{BHF1}.
Moreover, a question critical for interpreting all SRC experiments has been the effects of the final state interactions (FSI) suffered by the outgoing nucleons. As normally the SRC effects are on the order of 20\%, even a small FSI effect may influence what one can learn about the SRC itself. Compared to the reactions induced by a single nucleon, electron and photon, collisions between two heavy nuclei can make good uses of
the two HMTs existing in both the target and projectile. More specifically, the CMS energy of two colliding nucleons from the two HMTs of the target and projectile respectively will be much higher than those involving nucleons all from below the Fermi surfaces of the two colliding nuclei. These higher CMS energies available will make sub-threshold productions of various particles, such as high-energy photons, pions, kaons, nucleon-antinucleon pairs, etc, possible. In particular, the hard photon production via $pn\rightarrow pn\gamma$ process in heavy-ion collisions is expected to be increased by the SRC.
Since photons only interact with nucleons electromagnetically, it is the most FSI free probe of the reaction and may thus carry the most reliable information about the HMT in the initial target and projectile involved in the reaction.

Because of the strong isospin dependence of the HMT, it was already shown that the SRC affects the density dependence of $E_{sym}(\rho)$. For example, it has been shown that not only the kinetic part of the $E_{sym}(\rho)$ gets decrease with respect to the prediction of the free Fermi gas model in several independent studies \cite{CXu11,CXu13,Vid11,Lov11,car12,car14}, its high-density part also gets decreased \cite{caili2017,caili4}. Moreover, the SRC can lead to an appreciable isospin-quartic term \cite{liba15} in the EOS of neutron-rich matter, see, e.g., ref \cite{Li17review} for a recent review. The SRC-induced modifications of the $E_{sym}(\rho)$ may manifest themselves in several observables of heavy-ion reactions \cite{hen14,Li15,Yong1}.  In fact, because of its fundamental nature and broad impacts, nuclear symmetry energy has been extensively studied by both the nuclear physics and astrophysics communities, see, e.g., refs.~\cite{ireview98,ibook01,Steiner05,ditoro,LCK08,Trau12,Tsang12,Lat13,Hor14,LiBA14,Heb15,Bal16,Oer17,LiBA17}
for reviews. The study of nuclear symmetry energy is also one of the major scientific motivations of several recent experiments at several facilities, see, e.g., refs. \cite{frib,fair,exp1,exp2,koria,csr,TAMU1,TAMU2,TAMU3,Catania,Chuck}. It is thus important to find experimental observables sensitive to the $E_{sym}(\rho)$. Interestingly, hard photons have been identified as among the useful probes of the density dependence of $E_{sym}(\rho)$ and SRC \cite{yongp1,yongp2,Xue,yongp4}.

Given the fact that both the symmetry energy and the SRC in neutron-rich matter are not completely understood and they are important for resolving many interesting issues, it is necessary to explore the interplay of the SRC and symmetry energy on hard photon production in heavy-ion collisions. Hopefully our theoretical results will help shed new light on both topics. In fact, there are also interests by several experimentalists to explore both the
$E_{sym}(\rho)$ and SRC using hard photons. To make our calculations useful for the planned experiments at Texas A\&M University \cite{Alan}, in this work we study hard photon productions in several reactions involving Ca and Sn isotopes at 45 MeV/nucleon.

\section{The brief description of the updated IBUU transport model}

\begin{figure}[th]
\centering
\includegraphics[width=0.5\textwidth]{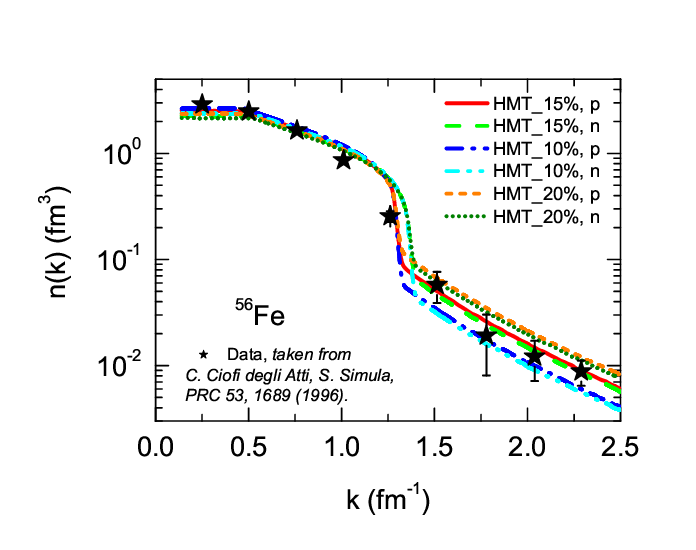}
\caption{ (Color online) Nucleon momentum distribution n(k) of $^{56}_{26}$Fe with normalization condition $\int_{0}^{\lambda k_{F}}n(k)k^{2}dk$ = 1.} \label{fe56}
\end{figure}
\begin{figure}[th]
\centering
\includegraphics[width=0.5\textwidth]{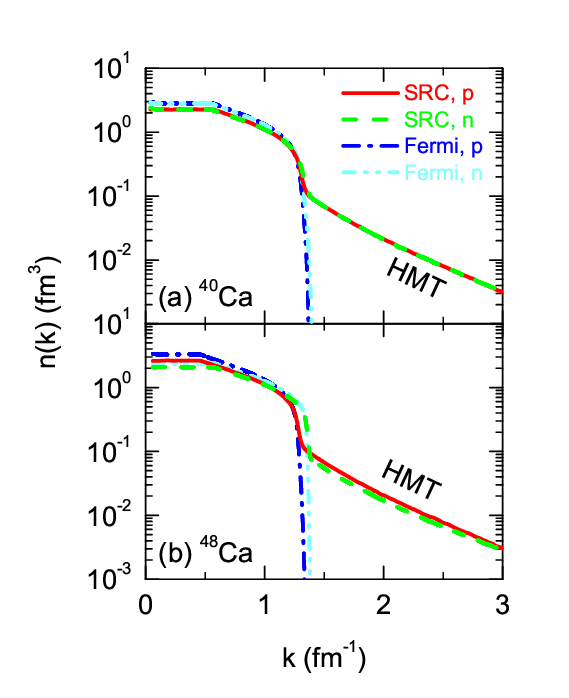}
\caption{ (Color online) Nucleon momentum distribution n(k) in $^{40}_{20}$Ca and $^{48}_{20}$Ca. For a comparison, the nucleon Fermi distribution is also shown.} \label{npdis}
\end{figure}

In the following, we shall first briefly describe how we incorporate the HMT in initializing nucleons in phase space and the main inputs of an Isospin-dependent
Boltzmann-Uehling-Uhlenbeck (IBUU) transport model used in this work. We initialize neutrons and protons using their density profiles predicted by the Skyrme-Hartree-Fock calculations using the Skyrme
M$^{\ast}$ force parameters \cite{skyrme86}. We divide each nucleus into many spherical shells centered around its CMS.  By using the local Thomas-Fermi approximation in each shell of radius $r$, the local Fermi momenta of neutrons and protons in each shell are calculated according to
\begin{equation}
 k_{F_{n,p}}(r)= [3\pi^{2}\hbar^{3}\rho(r)_{n,p}]^{\frac{1}{3}}.
\end{equation}
In each shell, the nucleon momenta are generated according to the following distributions with HMTs reaching $\lambda k_{F_{n,p}}(r)= 2.75\times k_{F_{n,p}}(r)$ \cite{henprc15}
\begin{eqnarray}
n(k)=\left\{%
  \begin{array}{ll}
    C_{1}, & \hbox{$k \leq k_{F}$;} \\
    C_{2}/k^{4}, & \hbox{$k_{F} < k < \lambda k_{F}$}\\
\end{array}%
\right.
\label{nk}
\end{eqnarray}
where $C_1$ and $C_2$ are constants determined by the total numbers as well as the specified fractions of neutrons and protons in their respective HMTs.
They are normalized by the condition
\begin{equation}
\int_{0}^{\lambda k_{F}}n(k)k^{2}dk = 1.
\label{nk1}
\end{equation}

Since for medium and heavy nuclei about 20\% nucleons \cite{e93,e96,sci08} are in the HMT, neglecting the detailed mass dependence and adopting the n-p dominance model requiring equal numbers of neutrons and protons in the HMT \cite{sci14}, 10\% of the total nucleons are assigned equally as neutrons or protons and distributed in their respective HMTs. The rest of them are then distributed in their respective Fermi seas.
Thus, the nucleon momentum distribution in nuclei can be formally written as
\begin{equation}
 n_{n,p}(k)= \frac{1}{N,Z}\int _{0}^{r_{max}}d^{3}r\rho_{n,p}(r)\cdot n(k,k_{F_{n,p}}(r))
\end{equation}
with $N$ and $Z$ being the total numbers of neutrons and protons in a nucleus. While the momentum distribution function $n_{n,p}(k)$ is not experimentally directly measurable, it can be inferred from model analyses of some experimental
observables, such as cross sections of electron-nucleus scattering.
As an illustration of the generated nucleon momentum distributions,
shown in Fig.~\ref{fe56} are the generated nucleon momentum distributions in $^{56}_{26}$Fe in comparison with that extracted from analyzing some experimental data by Ciofi degli Atti et al. \cite{Ciofi96}.
It is clearly seen that our initialization can reproduce their results quite well. In the calculations, various HMT fractions are used. In Fig.~\ref{npdis}, we plot the nucleon momentum distribution of Ca isotopes with or without the HMT. Compared with the ideal gas case, for the neutron-rich nucleus $^{48}_{20}$Ca, protons have a larger probability than neutrons to have momenta greater than the nuclear Fermi momentum. This feature is a consequence of the n-p dominance model where equal numbers of neutrons and protons are required to be in the HMT.

\begin{figure}[th]
\centering
\includegraphics[width=0.5\textwidth]{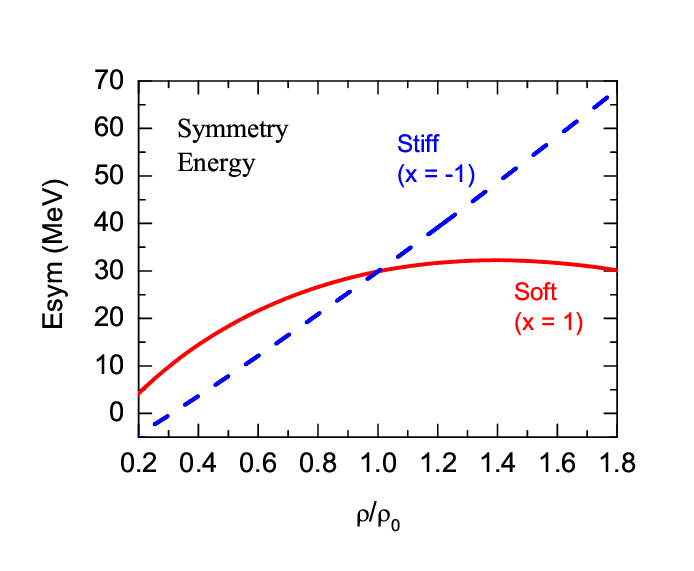}
\caption{ (Color online) The corresponding density-dependent symmetry energy in Eq. (\ref{buupotential}). x= -1, 1 are for the stiff and soft symmetry energies, respectively.} \label{desym}
\end{figure}
In this study, the following isospin- and momentum-dependent single-nucleon
potential (MDI) is used \cite{spp1,yongprc2016}
\begin{eqnarray}
U(\rho,\delta,\vec{p},\tau)&=&A_u(x)\frac{\rho_{\tau'}}{\rho_0}+A_l(x)\frac{\rho_{\tau}}{\rho_0}\nonumber\\
& &+B(\frac{\rho}{\rho_0})^{\sigma}(1-x\delta^2)-8x\tau\frac{B}{\sigma+1}\frac{\rho^{\sigma-1}}{\rho_0^\sigma}\delta\rho_{\tau'}\nonumber\\
& &+\frac{2C_{\tau,\tau}}{\rho_0}\int
d^3\,\vec{p^{'}}\frac{f_\tau(\vec{r},\vec{p^{'}})}{1+(\vec{p}-\vec{p^{'}})^2/\Lambda^2}\nonumber\\
& &+\frac{2C_{\tau,\tau'}}{\rho_0}\int
d^3\,\vec{p^{'}}\frac{f_{\tau'}(\vec{r},\vec{p^{'}})}{1+(\vec{p}-\vec{p^{'}})^2/\Lambda^2},
\label{buupotential}
\end{eqnarray}
where $\rho_0$ denotes saturation density, $\tau, \tau'=1/2(-1/2)$ for neutron (proton) and $\delta$ is the isospin asymmetry of the system.
Different symmetry energy's stiffness parameters $x$
can be used in the above single-nucleon potential to mimic
different forms of the symmetry energy predicted by various
many-body theories without changing any
property of the symmetric nuclear matter and the symmetry energy
at normal density. In this study, we choose x = 1 (default value) and -1 for the soft and stiff symmetry energies.
The corresponding density-dependent symmetry energy is shown in Fig.~\ref{desym}. By design, around the saturation density $\rho_{0}$, the soft symmetry energy has a small slope ($L (\rho_{0}) \equiv 3\rho_{0}dE_{sym}(\rho)/d\rho$) compared with the stiff symmetry energy. The stability of initial nuclei when HMT is incorporated was studied in Ref. \cite{yongprc2016} in the case of distributing all nucleons in a share sphere. It was found that about 5\% and 3\% nucleons are artificially emitted by 10 fm/c. In the present study using the SHF predicted density profiles for neutrons and protons, only nucleons in the interior have relatively high local Fermi momentum. We thus expect to
have less spurious nucleon emissions compared to that found in Ref. \cite{yongprc2016}. As we are focusing on hard photons mostly from the first chance neutron-proton scatterings, the level of instability does not pose a serious problem. For nucleon-nucleon collisions, the isospin-dependent reduced nucleon-nucleon scattering cross sections in medium are used \cite{yongprc2016,Lichen05}.

\begin{figure}[th]
\centering
\includegraphics[width=0.5\textwidth]{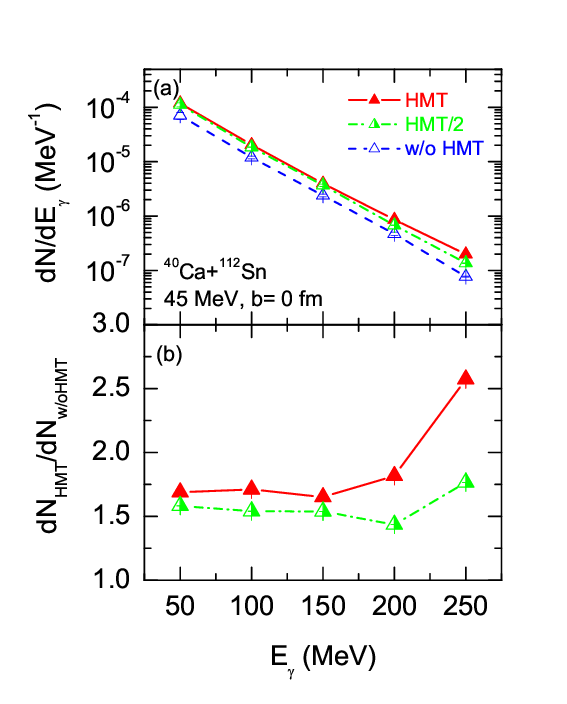}
\caption{ (Color online) Top panel: HMT effects on hard photon productions in head-on $^{40}$Ca + $^{112}$Sn reactions at 45 MeV/nucleon with the default value of x = 1. Bottom panel: The ratio of hard photon spectra with the HMTs and without HMT.} \label{hmtp}
\end{figure}
\begin{figure}[th]
\centering
\includegraphics[width=0.5\textwidth]{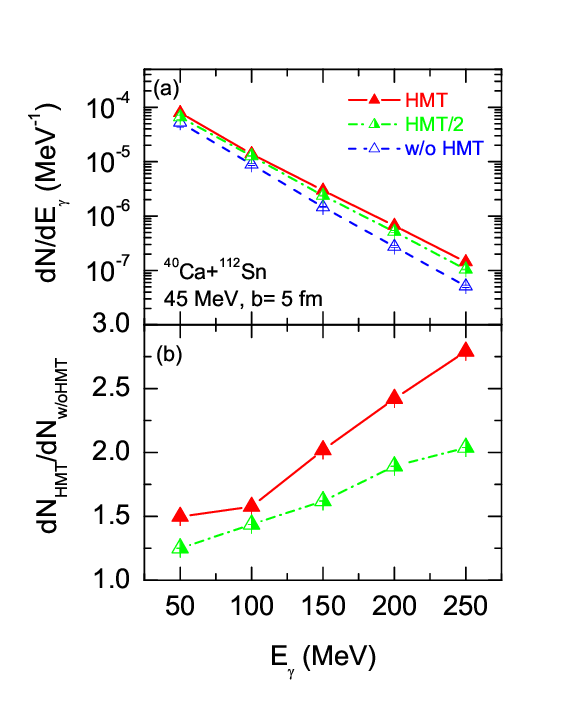}
\caption{ (Color online) Same as Fig.~\ref{hmtp}, but with an impact parameter of b= 5 fm.} \label{hmtp5}
\end{figure}

In fact, hard photons in heavy-ion collisions at intermediate energies have been studied in a number of previous works \cite{bertsch,grosse86,nif90,cassrp,yongp1,yongp2,yongp4}.
For example, the TAPS collaboration has done a series of experiments investigating properties of hot and dense matter
using hard photons as messengers \cite{TAPS,TAPS2,TAPS3,TAPS4}. Theoretically, it was concluded that neutron-proton bremsstrahlungs in the early stage of the reaction
are the main source of high energy photons \cite{may1,may2}.
It was also demonstrated that hard photon productions in heavy-ion collisions can be used to
probe the reaction dynamics leading to the formation of dense
matter \cite{bertsch86,ko85,cassing86,bau86,stev86}. Although the
elementary cross section for the $pn\rightarrow pn\gamma$ process used in transport simulations is still model
dependent \cite{nif85,nak86,sch89,gan94,tim06,saf07}, the old experimental data can be
described reasonably well \cite{cassrp} by using the following parameterization for the probability of hard photon production based on the one boson exchange model
\cite{gan94,yongp1,yongp2}
\begin{equation}\label{QFT}
p_{\gamma}\equiv\frac{dN}{d\varepsilon_{\gamma}}=2.1\times10^{-6}\frac{(1-y^{2})^{\alpha}}{y}.
\end{equation}
In the above, $y = \varepsilon_{\gamma}/E_{max}$, $\alpha = 0.7319-0.5898\beta_i$,
$\varepsilon_{\gamma}$ is the energy of photon emitted, $E_{max}$ is the energy
available and $\beta_i$ is the initial velocity of
the proton in the colliding proton-neutron CMS frame.
The Pauli-blockings of final state nucleons in the $pn\rightarrow pn\gamma$ process are also taken into account as in Ref. \cite{bau86}.

\section{Results and discussions}

We focus on investigating the interplay of SRC and $E_{sym}(\rho)$ effects on hard photon productions. Unless specified otherwise,  we shall present results obtained by using a 20\% HMT fraction.
The upper panel of Fig.~\ref{hmtp} shows effects of the HMT on the hard photon spectrum. It is clearly seen that with the HMT, there is a clear increase of hard photon production. This is because the hard photons, especially the most energetic ones, are mainly from neutron-proton collisions with larger CMS energies involving nucleons
from the two HMTs in the target and projectile. To demonstrate more quantitatively effects of the HMT, we also did a calculation with a 10\% HMT fraction. From the lower panel of Fig.~\ref{hmtp}, it is seen that with the HMT halved, the probability of producing photons with energies around 250 MeV is reduced by a factor about 0.7.
Similar results are shown for the case with a larger impact parameter of 5 fm in Fig.~\ref{hmtp5}.
For photons with energies significantly less than 250 MeV, they are from collisions involving nucleons from either one HMT and one Fermi sea or two Fermi seas.
Comparing results in Fig.~\ref{hmtp} and Fig.~\ref{hmtp5}, it is seen that hard photon production in calculations including HMT effects is very similar with impact parameters b = 0 and b = 5 fm.
Our results indicate that nucleon-nucleon collisions involving HMT nucleons play a major role in producing hard photons with energies above 50 MeV.

\begin{figure}[th]
\centering
\includegraphics[width=0.5\textwidth]{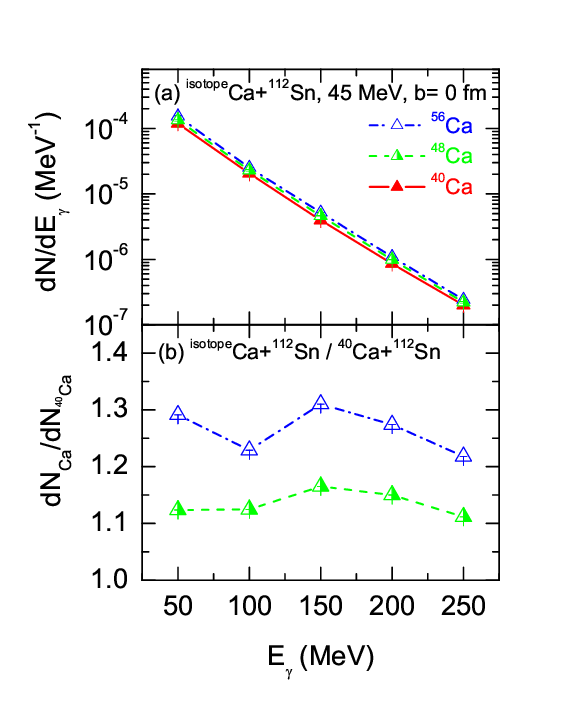}
\caption{ (Color online) Top panel: Hard photon productions in head-on collisions of Ca isotopes on $^{112}$Sn target at 45 MeV/nucleon with the default value of x = 1. Bottom panel: The ratio of hard photon productions in reactions with $^{56,48}$Ca projectiles over that with $^{40}$Ca.} \label{iso}
\end{figure}
\begin{figure}[th]
\centering
\includegraphics[width=0.5\textwidth]{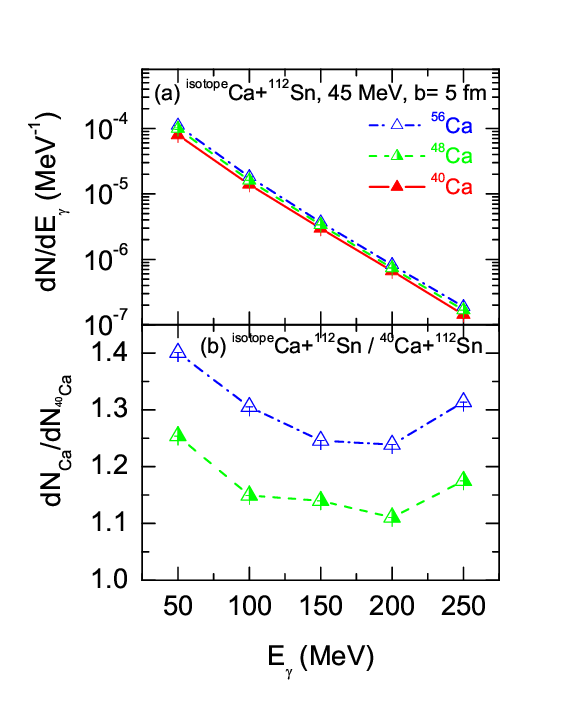}
\caption{ (Color online) Same as Fig.~\ref{iso}, but with an impact parameter of b= 5 fm.} \label{iso5}
\end{figure}
Next, we study the mass dependence of hard photon production by comparing results of reactions induced by three Ca isotopes on the same $^{112}$Sn target at 45 MeV/nucleon with impact parameters of 0 and 5 fm, in
Fig.~\ref{iso} and Fig.~\ref{iso5},  respectively. It is seen that the energy spectra of hard photons with different Ca isotopes are quite similar at both impact parameters. One can also examine the ratios of hard photon spectra in reactions induced by $^{56}$Ca, $^{48}$Ca over $^{40}$Ca. As expected,  more photons are produced in the heavier and more neutron-rich reaction in the whole energy range considered.
More quantitatively, in central collisions the hard photon production with $^{48}$Ca increases by a factor of about 1.125 compared to that using the $^{40}$Ca projectile. While with $^{56}$Ca, the hard photon production increases by a factor of about 1.25. The more production of hard photons with increasing neutron number in the projectile is simply due to the more abundant neutron-proton collisions happened.

\begin{figure}[th]
\centering
\includegraphics[width=0.5\textwidth]{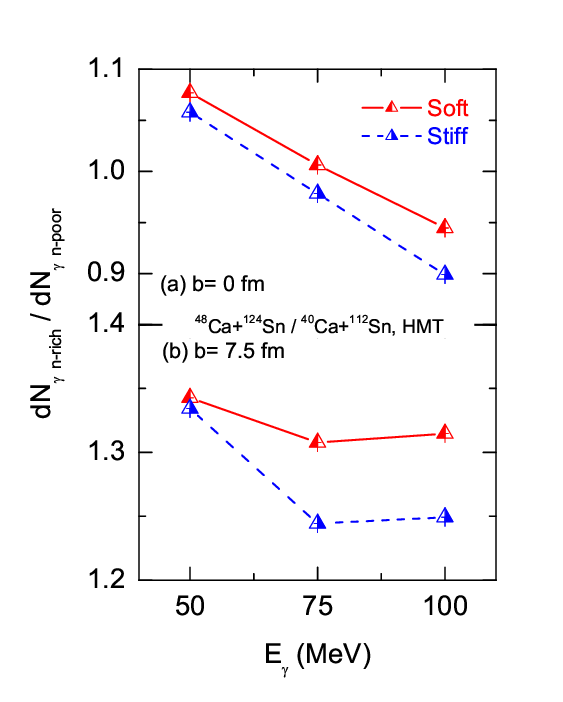}
\caption{ (Color online) Effects of the symmetry energy on the ratio of hard photon productions in neutron-rich and neutron-deficient reactions at a beam energy of 45 MeV/nucleon with different impact parameters.} \label{esym}
\end{figure}
Since the study of nuclear symmetry energy is of great importance in both nuclear physics and astrophysics, significant efforts have been made by many people to constrain its
density dependence that is still poorly known especially at supra-saturation densities. One of the major challenges is that the symmetry energy term $E_{sym}(\rho)\cdot \delta^2$ is relatively small compared to the symmetric part of the EOS under conditions reachable in terrestrial nuclear reactions. Generally speaking, symmetry energy effects are thus all rather small. Moreover, if one uses strong-interacting particles as messengers, the already rather weak signal of symmetry energy may get distorted by the FSI. Therefore, it is advantageous to have clean probes free of the FSI. Indeed, it was first shown in Ref.~\cite{yongp1} that the hard photon production in heavy-ion collisions is a promising one in this respect. Moreover, it was found that the soft symmetry energy leads to more hard photon productions compared to the stiff symmetry energy in heavy-ion reactions at beam energies around 50 MeV/nucleon. However, the SRC effects were not considered at the time. It is thus useful to know whether the symmetry energy still affects the hard photon production when the SRC effects are also considered in the same calculation. Moreover, at what level can one learn anything about either or both the SRC and  $E_{sym}(\rho)$ if at all possible from hard photons in heavy-ion reactions?
Trying to answer these questions, we show in Fig.~\ref{esym} effects of the symmetry energy on the ratio of hard photons in neutron-rich ($^{48}$Ca+$^{124}$Sn) over neutron-poor $^{40}$Ca+$^{112}$Sn reactions at an incident energy of 45 MeV/nucleon with a 20\% HMT using the soft and stiff $E_{sym}(\rho)$ functions, respectively.
From the upper and lower panels of Fig.~\ref{esym}, it is clearly seen that the symmetry energy affects appreciably hard photon production in both the central and peripheral collisions. More quantitatively,  about 5\% more photons are produced at 100 MeV with the soft symmetry energy compared to that with the stiff $E_{sym}(\rho)$. This observation is qualitatively consistent with that obtained in Ref.~\cite{yongp1,ma2012}.
Moreover, for peripheral collisions of neutron-rich systems, neutron-skins make the isospin asymmetry larger and the number of neutron-proton collisions smaller as well as the SRC weaker, effects of the symmetry energy on photons are thus stronger. Considering related studies in Ref.~\cite{yongp1}, in the present study, the ratio of hard photon from the two reaction systems still mainly probes the symmetry energy at densities above nuclear saturation density.
As we discussed earlier, energetic photons are mostly from colliding neutrons and protons in the HMTs (where the isospin asymmetry is approximately zero) in the target and projectile. The $E_{sym}(\rho)$ has little effect on very energetic photons. We thus show in Fig.~\ref{esym} only the hard photons with energies up to 100 MeV.

We now turn to the interplay between the SRC and $E_{sym}(\rho)$ effects on hard photons. First of all, it is worth emphasizing that effects of the $E_{sym}(\rho)$ on hard photons depend not only on the system size but also the exact fraction of nucleons in the HMT. This is because for specific colliding nuclei at a given beam energy and impact parameter, different numbers of neutron-proton pairs in the HMT would lead to different isospin-asymmetries below and above the Fermi momentum. Because of the isospin- and momentum-dependent single-nucleon potential, the variation of nucleon isospin-asymmetry below and above the Fermi momentum then influences the effects of the $E_{sym}(\rho)$ on the hard photon spectra as we analyzed above. On the other hand, as we have shown in Figs.~\ref{hmtp} and \ref{hmtp5}, for hard photons with energies below about 150 MeV, their yield ratio does not change much by varying the HMT fraction from 10\% to 20\%. However, the ratio of hard photons from calculations with over without the HMT ranges from about 50\% in the peripheral collision to a factor of 3 in head-on collisions.
Thus, in the whole energy range of photons, effects of the SRC overwhelm those due to the  $E_{sym}(\rho)$. Moreover, for the most energetic photons, they are all from the HMTs where the nucleon isospin-asymmetry is about zero. Thus, the high energy photons can be used to probe properties of the HMT with little influence from the $E_{sym}(\rho)$. As we discussed earlier, there are many interesting issues regarding the size, shape and isospin dependence of the HMT. Our findings here indicate that the hard photons from heavy-ion collisions provide a much more clean means to probe the HMT free of the FSI of outgoing nucleons in $e-A$ and $p-A$ reactions. Obviously, until the HMT is better understood, it is practically impossible to constrain the $E_{sym}(\rho)$ using hard photons from heavy-ion collisions. We notice that the neutron-proton effective mass splitting associated with the momentum-dependent isovector potential is another factor that is still uncertain and may have some effects on hard photon production through both the elementary $pn\rightarrow pn\gamma$ cross section and the reaction dynamics. However, effects of the former are largely cancelled out in the ratios we examined in this work. While effects of the neutron-proton effective mass splitting on the reaction dynamics and observables are at most at the same level as the $E_{sym}(\rho)$ \cite{balinpa2003}. We thus expect the SRC effects to remain dominant on hard photon production. Nevertheless, since the elementary $pn\rightarrow pn\gamma$ cross section is determined by the neutron-proton relative velocity, once SRC effects are well understood or if one can find proper observables in single reaction systems, it would be interesting to investigate if photons can help us extract information about nucleon effective masses in dense nuclear matter.

It is also worth noting that the ratio of energetic photons from the neutron-rich system over neutron-deficient system becomes appreciably larger with increasing impact parameter as shown in Figs.~\ref{iso}, \ref{iso5} and \ref{esym}.
This is mainly because we are taking the ratios of two reactions with different masses. At the same impact parameter, the participant region of the lighter and neutron-poor system is smaller. Thus, as the impact parameter increases from b=0 to 5fm, the yield of energetic photons from the lighter neutron-deficient system decreases more compared with that of the neutron-rich reaction studied, leading to the higher ratios at b=5 fm.

\section{Summary}
The new physics underlying both the short-range correlations and symmetry energy in neutron-rich matter is fundamentally important for both nuclear physics and astrophysics.
The physics ingredients of the SRC and $E_{sym}(\rho)$ are actually closely intercorrelated. Significant efforts have been made by many people to probe both the SRC and $E_{sym}(\rho)$ using various theoretical approaches and experimental methods. Among the promising probes known, hard photons from heavy-ion collisions have the special advantages that it is basically free of the final state interactions that have been the major sources of uncertainties in interpreting some experimental findings from studying hadronic probes. However, so far not much research has been done about the interplay of the SRC and $E_{sym}(\rho)$ effects on hard photon productions in heavy-ion collisions at low and intermediate energies neither experimentally nor theoretically.

Motivated by the fundamental importance of better understanding both the SRC and $E_{sym}(\rho)$ as well as the strong interest of some experimental groups to actually measure hard photons,  we investigated the interplay of the
SRC and $E_{sym}(\rho)$ effects on hard photon spectra in collisions of several Ca isotopes on $^{112}$Sn and $^{124}$Sn targets at a beam energy of 45 MeV/nucleon.
We found that over the whole energy range of hard photons considered, effects of the SRC overwhelm those due to the  $E_{sym}(\rho)$. The energetic photons come mostly from the high-momentum tails where the nucleon isospin-asymmetry is zero within the neutron-proton dominance model of SRC. These high-energy photons are very sensitive to the features of the high-momentum tails with little influence from the $E_{sym}(\rho)$. Therefore, experimental measurements of the energetic photons from heavy-ion collisions at Fermi energies have the potential to help us better understand the nature of SRC. These measurements will be complementary to but also have some advantages over the ongoing and planned experiments using hadronic messengers from reactions induced by high-energy electrons or protons. It is known that the SRC  reduces (enhances) the kinetic (potential) contribution to the $E_{sym}(\rho)$ and also increase the isospin-quartic term in the EOS of neutron-rich matter. The SRC also makes the $E_{sym}(\rho)$ more concave around the saturation density of nuclear matter, thus affecting the isospin dependence of nuclear incompressibility currently being studied by measuring various giant resonances along isotope chains. Thus, a better understanding of the SRC will also improve our knowledge about the EOS especially the $E_{sym}(\rho)$ in neutron-rich matter. We are enthusiastically looking forward to comparing our calculations with the forthcoming experimental data.

\section*{Acknowledgements}
We would like to thank Alan B. McIntosh for communications and discussions. This work is supported in part by the National Natural Science
Foundation of China under Grant Nos. 11375239, 11775275, 11435014, 11320101004 and the U.S. Department of Energy under Award Number DE-SC0013702 and the CUSTIPEN (China-U.S. Theory Institute for Physics with Exotic Nuclei) under the US Department of Energy Grant No. DE-SC0009971.

\end{document}